\documentstyle[fleqn] {revp}
\newfont{\sff}{cmssi12} 
\newfont{\bigsf}{cmss12 scaled 2000} 
\newfont{\midsf}{cmss12 scaled 1000} 
\newfont{\smlsf}{cmss12 scaled 600}  
\newfont{\bigsff}{cmssi12 scaled 2000} 
\newfont{\sfi}{cmssi10} 

\newcommand{\pmonth}{March, 1999}
\newcommand{\spec}{Selected Topics in Nuclear Collective Excitations 
(NUCOLEX99)}
\newcommand{\no}{10}

\begin{document}
\parindent 0pt
\parskip 12pt
\setcounter{page}{1}

\title{Compressibility of Nuclear Matter from Shell Effects in Nuclei}

\author{M.M. Sharma\\
\sff
Physics Department, Kuwait University, Kuwait 13060}

\abst{
The compressibility of nuclear matter has received significant attention
in the last decade and a variety of approaches have been employed to
extract this fundamental property of matter. Recently, significant 
differences have emerged between the results of relativistic and 
non-relativistic calculations of breathing mode giant monopole 
resonance (GMR). This is due to a lack of understanding of the
dynamics of GMR and of its exact relationship to the compression 
modulus of the infinite nuclear matter. Here, I present
an alternative approach based upon nuclear shell effects.
The shell effects are known to manifest experimentally in terms of 
particle-separation energies with an exceedingly high precision. 
Within the framework of the non-relativistic density-dependent Skyrme 
theory, it will be shown that the compressibility of nuclear matter has a 
significant influence on shell effects in nuclei. I will also show that 
2-neutron separation energies and hence the empirical shell effects can 
be used to constrain the compressibility of nuclear matter.}

\maketitle
\thispagestyle{headings}

\section*{Introduction}
Shell effects manifest themselves in terms of the existence of magic
numbers in nuclei. Inclusion of spin-orbit coupling in the shell model
is known to account for the magic numbers$^{1)}$.
Shell effects are also found to appear more subtly 
e.g. as anomalous kink in the isotope shifts of Pb nuclei. 
It was shown that in the relativistic mean-field (RMF) theory, 
this kink appears naturally due to the inherent Dirac-Lorentz 
structure of nucleons$^{2)}$. By including an
appropriate isospin dependence of the spin-orbit potential in the
density dependent Skyrme approach, the anomalous kink in the isotope 
shifts in Pb nuclei could be produced$^{3)}$.

The shell gaps at the magic numbers are known to produce important effects.
Consequences due to the shell gaps are broadly termed as shell effects. 
The clearest manifestation of the shell effects can be seen in 
the pronounced dip in neutron and proton separation energies at
the magic numbers in figures by Wapstra and Audi$^{4)}$ on
the experimental separation energies all over the periodic table.
Recently, shell effects have become focus of much attention$^{5,6)}$.
This is due to the reason that the shell effects in nuclei near the drip 
lines constitute an important ingredient to understand r-process abundances 
and heavy nucleosynthesis$^{7,8)}$. 

In this talk, I focus upon 2-neutron separation energies to
demonstrate that the shell efffects show a significant dependence
upon the compressibility of nuclear matter. Using empirical data
on the 2-neutron separation energies and thus implicitly the empirical
shell effects, I will show that these data can be used to constrain
the compressibility of nuclear matter. As the focus of this conference
shows, the compression modulus K is an important fundamental property
of the nuclear matter. It represents a cardinal point on the behaviour of 
equation of state (EOS) of the nuclear matter. 

The compressibility of the nuclear matter has received a significant attention 
in the last decade and various approaches$^{9-11)}$ have been employed 
to extract the compression modulus. However, an unambiguous 
determination of the compression modulus of the nuclear matter has remained 
difficult due to a lesser sensitivity of the giant monopole resonance
data to the compressibility$^{12,13)}$.

\section*{The Density-Dependent Skyrme Theory}

Here I employ the density-dependent Skyrme theory$^{14)}$ to examine 
the shell effects. The Skyrme approach has generally been found to
be very successful in providing ground-state properties of nuclei. 
The energy density functional used  in the Skyrme approach is the
standard one as given in Ref.$^{15)}$. The corresponding energy 
per nucleon for the symmetric nuclear matter (with the Coulomb force
switched off) is given by
\begin{equation}
(E/A)_\infty = k\rho^{2/3}(1 + \beta\rho) + {3\over8} t_0\rho
+ {1\over 16}t_3\rho^{1+\alpha}
\end{equation}
where $k = 75$ MeV.fm$^2$. The constant $\beta$ is given in terms of the
constants $t_1$ and $t_2$ and $x_2$ of the Skyrme force by
\begin{equation}
\beta = {2m\over\hbar^2}{1\over4}\Big[{1\over4}(3t_1 + 5t_2) +
t_2x_2\Big].
\end{equation}
The constant $\beta$ is also related to the effective mass $m^*$ and
the saturation density $\rho_0$ by
\begin{equation}
m/m^* = 1 +\beta\rho_0
\end{equation}
The incompressibility (or the compression modulus) of the infinite
nuclear matter is given as the curvature of the EOS curve and
can be written as 
\begin{eqnarray}
K &= &9\rho_0^2{d^2(E/A)(\rho)\over d\rho^2}\Big |_{\rho_0} \\  
\nonumber 
  &= &-2k\rho_0^{2/3} + 10k\beta\rho_0^{5/3} + 
{t_3\over 16}\alpha(\alpha + 1)\rho_0^{1 + \alpha}
\label{eqn}
\end{eqnarray}
The parameters $t_0$, $t_3$ and $\alpha$ are usually obtained from the
nuclear matter properties. The other parameters $t_1$, $t_2$ and
various $x$ parameters are obtained from fits to properties
of finite nuclei. The strength W$_0$ responsible for the spin-orbit 
interaction is obtained by reproducing the spin-orbit splittings in 
nuclei such as $^{16}$O and $^{40}$Ca. 

\section*{The Skyrme Forces}

In order to show the effect of the incompressibility of nuclear matter
on shell effects and consequently also on ground-state properties of nuclei, 
I have constructed a series of zero-range Skyrme forces. Experimental data
on ground-state binding energies and charge radii of key nuclei such
as $^{16}$O, $^{40}$Ca, $^{90}$Zr, $^{116}$Sn, $^{124}$Sn and $^{208}$Pb  
are taken into account in the least-square minimization. The ground-state
properties in the Skyrme theory are calculated using the Hartree-Fock method.

With a view to vary K over a large range, the correlation between 
K and the saturation density is implicitly taken into account. 
This correlation has been summarized for Skyrme type of 
forces in Fig. 4 of Blaizot$^{9)}$. Accordingly, there exists an 
inverse correlation between the saturation density (Fermi momentum)
and the compression modulus. Keeping this in mind, I have varied the 
saturation density $\rho_0$ over the range $0.140 - 170 fm^{-3}$ 
in steps of $.005 fm^{-3}$ as shown in Table 1. 

The effective mass is fixed at 0.79. This value is 
required for being consistent with giant quadrupole resonance
energies of heavy nuclei$^{15)}$. Eq. (3) then provides
the value of $\beta$ which can be used in eq. (2) to connect
the coefficients $t_1$, $t_2$ and $x_2$. In order to do a systematic 
variation of the nuclear compressibility, I have kept the saturation 
binding energy fixed at a value of $-16.0$ MeV. This value is consistent 
with most nuclear mass models$^{16)}$ and is close to physically 
acceptable values. 

The coefficients $\alpha$, $t_0$ and $t_3$ are determined by the
eqs. (1), (4) and the saturation condition. However, we have 
allowed $t_0$ and $t_3$ to vary so that a fit to finite nuclei
provides an incompressibility that is consistent with the inverse 
correlation.  On the other hand, in an exhaustive computational 
exercise it is found that the violation of the above correlation 
results in bad fits for ground-state binding energies of key nuclei. 

I have fixed $x_1$ and $x_2$ at zero for convenience. Given a value 
of $\beta$, the coeffients $t_1$ and $t_2$ in eq. (2) are obtained 
from the fits to the ground-state binding energies of finite nuclei.
Thus, I have obtained various Skyrme parameter sets with 
incompressibility values K = 200, 220, 249, 270, 305, 327, 360 and 
393 MeV, respectively. These forces encompass a broad range of physically 
plausible values of K. The nuclear matter properties of these forces 
are shown in Table 1. It can be seen that the force with K = 270 MeV
was obtained for the sake of interpolation. Table 2 shows the total
binding energies of some key nuclei obtained with the HF+BCS approach
using the various Skyrme forces. It can be seen that all of these forces 
reproduce the total binding energies of the key nuclei from $^{16}$O to 
$^{208}$Pb very well.

\section*{Two-neutron Separation Energies and the Compressibility}

I have selected the chain of Ni isotopes in order to probe the shell
effects. As most Ni isotopes are known to be spherical and the experimental 
binding energies are known over a large range of Ni isotopes, 
the chain of Ni isotopes serves as an ideal test bench to probe the 
shell effects. Even-mass Ni isotopes from A=52 to A=70 are considered. 
This includes the neutron magic number N=28 (A=56) where we intend 
to investigate the shell effects due to the major shell closure.

With a view to obtain the ground-state properties of nuclei,
we have performed spherical Skyrme Hartree-Fock calculations in
coordinate space. Herein pairing is included using the BCS formalism
with constant pairing gaps. Since nuclei under focus such as $^{56}$Ni
and $^{58}$Ni are not far away from the stability line, the BCS
scheme suffices to be a suitable mechanism for the pairing. 
Ground-state binding energies and $rms$ charge and neutron radii for 
the Ni isotopes are calculated using the various Skyrme forces. 
The binding energies of nuclei are used to obtain the 2-neutron 
separation energies as
\begin{equation}
S_{2n}(Z,N) = B (Z, N) - B (Z, N-2),
\end{equation}
where B represents the total binding energy of a nucleus.

As mentioned earlier, effects due to the shell gaps affect various
nuclear properties besides the particle separation energies.
First, the calculated charge radii of Ni isotopes are shown in Fig. 1.
The charge radii show an increase with the incompressibility.
Evidently, forces with the incompressibility about 200-250 MeV
show a little dispersion in the values of the charge radii. However,
above $K \sim 300$ MeV, there is a clear increase in the charge radii as
a function of K and forces with a large incompressibility
yield large values of charge radii for any given isotope.
Inevitably, a large value of K hinders synthesis of nuclei with a 
radial extension which is diminuted (compressed) contrary to 
that with a lower values of K. The empirical $rms$ charge radii of 
$^{58}$Ni, $^{60}$Ni, $^{62}$Ni and $^{64}$Ni are taken from the 
compilation of Ref.$^{17)}$ by solid circles. The values are 
3.776, 3.815, 3.846 and 3.868 fm, respectively, obtained 
by folding the proton density distributions with the finite 
size 0.80 fm of protons. These results derive from experiments on
muonic atoms. The $rms$ charge radii deduced from the precision 
measurements in muonic atoms (taken from the recent compilation of 
Ref.$^{18)}$ are 3.776, 3.813, 3.842 and 3.860 fm, respectively, 
which are almost identical to the previous compilation$^{17)}$. 

The empirical charge radii shown in Fig. 1 encompass the theoretical
curves between K=270-327 MeV.  Whereas the charge radius for $^{58}$Ni
points towards K=270 MeV, that for the heavier Ni isotopes
is in the vicinity of K=327 MeV. Since the charge radii deduced from 
various methods have been obtained with a significant precision, 
these data could, in principle, be used to constrain the incompressibility. 
However, as there is still some model dependence in extraction of the
charge radii, I would like to withhold any conclusions from this figure.
On the other hand, the general trend of the experimental data points
to a higher value of the compression modulus.

Figure 2 shows the corresponding $rms$ neutron radii. The neutron
radii show a monotonous increase with the mass number, with the
exception of a slight kink at the magic number N=28. The change in the
$r_n$ values with K is similar to that for the charge radii, i.e. 
for lower values of K, there is a very little change in $r_n$ with K. 
A significant change in the neutron radii, however,  can be seen with large
K values. The empirical $rms$ neutron radii obtained from 800-MeV 
polarized-proton scattering experiment$^{17,19)}$ for the 
isotopes $^{58}$Ni and $^{64}$Ni are shown in the figure by the solid 
points. The values lie between the curves for K=300 and K=327 MeV. 

The 2-neutron separation energy $S_{2n}$ for the Ni isotopes obtained 
with the various Skyrme forces are shown in Fig. 3. For the sake of
clarity of the presentation, I have selected a set of the Skyrme
forces in the figure. The dramatic fall in the $S_{2n}$ curves is seen
conspicuously for the nucleus just above the N=28 magic number.  
Such a kink in the $S_{2n}$ values signifies the presence of the shell
effects which arise from closure of a major shell. All the Skyrme
forces produce such a kink in Fig. 3. However, the slope of the kink
between A=56 (N=28) and A=58 (N=30) changes from one force to the
other (as a function of the incompressibility $K$).
The difference in the $S_{2n}$ values of $^{56}$Ni and $^{58}$Ni
can be taken as a measure of the shell effects. The force
with the lowest incompressibility ($K=200$ MeV) gives this
difference as 6.59 MeV with a slope which is minimum amongst
all the curves. This implies that the shell effects due to this
force are the weakest one. As the K value increases, the corresponding 
difference shows a smooth increase and the ensuing steepness increases 
gradually. For the force with the highest incompressibility ($K=393$ MeV), 
the difference amounts to 9.46 MeV. This difference is higher than the 
experimental difference$^{20)}$ of 8.37 MeV. It implies that the shell 
effects show a strong dependence on the compression modulus K.
Notwithstanding this correlation, the difference in the experimental values 
of $S_{2n}$ can serve as a calibration for K. Consequently, the experimental 
difference and the slope is closest to the curve for K=270 MeV. 
Thus, on the basis of the empirical $S_{2n}$ values, I infer that the 
incompressibility of nuclear matter lies in the neighbourhood 
of $K \sim 270$ MeV.

\section*{Shell Effects at the Neutron Drip-Line}

Shell effects near the drip lines have been a matter of discussion
of late$^{5,6)}$. It has been shown earlier that within the
non-relativistic approaches of the Skyrme type, the shell effects near 
the drip lines and in particular near the neutron drip line are 
quenched$^{5)}$. This is contrary to that shown in the RMF theory 
by Sharma et al.$^{6)}$, where the shell effects were observed to
remain strong. Before such a debate is resolved, it is important to 
settle the issue of the shell effects near the stability line. 
Fig. 4 shows the 2-neutron separation energies for the Ni isotopes 
using the force SkP within the Hartree-Fock Bogolieubov approach$^{21)}$.
The force SkP has been found to be successful in reproducing the 
ground-state properties of nuclei. The comparison of the SkP results 
with the experimental $S_{2n}$ values shows that empirically there 
exist strong shell effects than predicted by the Skyrme force SkP. 
The compression modulus of the force
SkP is about 200 MeV. Thus, the behaviour of the force SkP about the
weak shell effects is consistent with the results from forces with
low value of the incompressibility in Fig. 3. This suggests again
that the shell effects in Ni are commensurate to a higher value of 
the compression modulus of the nuclear matter as inferred above.

How the shell effects behave along the neutron drip line as a function of
the compression modulus can be visualized in Fig. 5. Here I have chosen 
the chain of Zr isotopes. This chain includes nuclei on both the sides of 
the magic number N=82 (A=122). The total binding energies obtained in the
Hartree-Fock calculations with the various Skyrme forces are shown. For 
nuclei from A=116 (N=76) to A=122 (N=82), the total binding energy shows a 
steep increase in the value with all the forces. The forces with lower
K show a stronger binding in general as compared to those with higher
K, which is as expected. For nuclei above A=122, there is a striking 
difference in the way the binding energies progress with mass number.
For the force with K = 200 MeV, the binding energy of nuclei heavier 
than $^{122}$Zr shows an increase in the value with the mass number,
implying that a further sequential addition of a pair of neutrons to
the N=82 core does contribute to the binding energy.  As the K value
increases, the binding energy contribution from neutrons above N=82
starts diminishing. For K values 300 MeV and above, a stagnation
is observed in the binding energies. Such a behaviour implies
that the shell effects become stronger as the incompressibility
increases. This is again an indication that the shell effects are
strongly correlated to the incompressibility of the nuclear matter.
Consequently, the incompressibility inferred from the ground-state 
data along the stability line suggests that the shell effects near 
the neutron drip line are strong. This is consistent with the 
strong shell effects about the neutron drip line as concluded in 
the RMF theory$^{6)}$. 

For nuclei near the drip lines, the HFB approach is considered to
be more suitable than the HF+BCS one. However, the effects of the 
continuum are expected to be significant more for nuclei very 
near the drip line and in particular the observables which are 
most affected by coupling to the continuum are the $rms$ radii. 
The total binding energies are known to show a little difference 
in the HF+BCS and HFB approaches. The main character of the shell 
effects is not altered by the inclusion of the continuum, 
as has been pointed out in Ref.$^{8)}$ using the Skyrme force SIII.  
However, it will be interesting to see how a judicious choice of a
force compatible with the experimental data as discussed above, will
affect the r-process nucleosynthesis.

\section*{Summary and Conclusions}

In the present work, I have shown that there exists a correlation between
the shell effects and the compressibility of nuclear matter. The latter
is shown to influence the shell effects significantly. Consequently, 
the 2-neutron separation energies show a strong dependence on
the compression modulus. The ensuing correlation then provides a 
calibration for the compression modulus of nuclear matter. It is shown 
that using this correlation, the empirical data on the 2-neutron
separation energies and thus implicitly the shell effects can be used to 
constrain the compressibility of nuclear matter. This procedure allows 
me to conclude that the incompressibility K should be within 270-300 MeV. 
This conclusion is also consistent with the experimental data on 
charge radii. A natural consequence of the present study is that the 
shell effects near the neutron drip line are predicted to be strong. 
This is in conformity with the strong shell effects observed at
the neutron drip line in the RMF theory.

\section*{References}

\re
1) M.G. Mayer and J.H.D. Jensen, {\it Elementary Theory of 
 Nuclear Shell Structure} (Wiley, New York, 1955).
\re
2) M.M. Sharma, G.A. Lalazissis and P. Ring, Phys. Lett.
{\bf 317}, 9 (1993). 
\re
3) M.M. Sharma, G.A. Lalazissis, J. K\"onig,
    and P. Ring, Phys. Rev. Lett. {\bf 74} 3744 (1994).
\re
4) C. Borcea, G. Audi, A.H. Wapstra and P. Favaron,
Nucl. Phys. {\bf A565} 158 (1993).
\re
5) J. Dobaczewski et al., Phys. Rev. Lett. {\bf 72} 981 (1994);
ibid {\bf 73} 1869 (1994).
\re
6) M.M. Sharma, G.A. Lalazissis, W. Hillebrandt and P. Ring,
Phys. Rev. Lett. {\bf 72} 1431 (1994); ibid {\bf 73} 1870 (1994).
\re
7)  K.-L. Kratz et al., Astrophys. J. {\bf 403} 216 (1993).
\re
8) B. Chen et al., Phys. Lett. {\bf B355} 37 (1995).
\re
9) J.P. Blaizot, Phys. Rep. {\bf 64} 171 (1980).
\re
10) M.M. Sharma et al., Phys. Rev. {\bf C38} 2562 (1988).
\re
11) M.V. Stoitsov, P. Ring and M.M. Sharma, Phys. Rev.
{\bf C50} 1445 (1994).
\re
12) J.M. Pearson, Phys. Lett. {\bf B271} 12 (1991).
\re
13) S. Shlomo and D.H. Youngblood, Phys. Rev. {\bf C47} 529 (1993).
\re
14) D. Vautherin and D.M. Brink, Phys. Rev. {\bf C7} 296 (1973).
\re
15) M. Brack, C. Guet and H.B. Hakansson, Phys. Rep. {\bf 123}
275 (1985).
\re
16) P. M\"oller, J.R. Nix, W.D. Myers and W.J. Swiatecki,
At. Data Nucl. Data Tables {\bf 59} 185 (1995).
\re
17) C.J. Batty, E. Friedmann, H.J. Gils and H. Rebel,
Adv. Nucl. Phys. {\bf 19} 1 (1989).
\re
18) G. Fricke et al., Atomic Data and Nuclear Data Tables
{\bf 60} 178 (1995).
\re
19) L. Ray, Phys. Rev. {\bf C19} 1855 (1979).
\re
20) A.H. Wapstra and G. Audi, Nucl. Phys. {\bf A565} 1 (1993).
\re
21) J. Dobaczewski et al., Nucl. Phys. {\bf A422} 103 (1984) and private
communication (1994).

\newpage

\leftline{\Large {\bf Figure Captions}}
\parindent = 2 true cm
\begin{description}

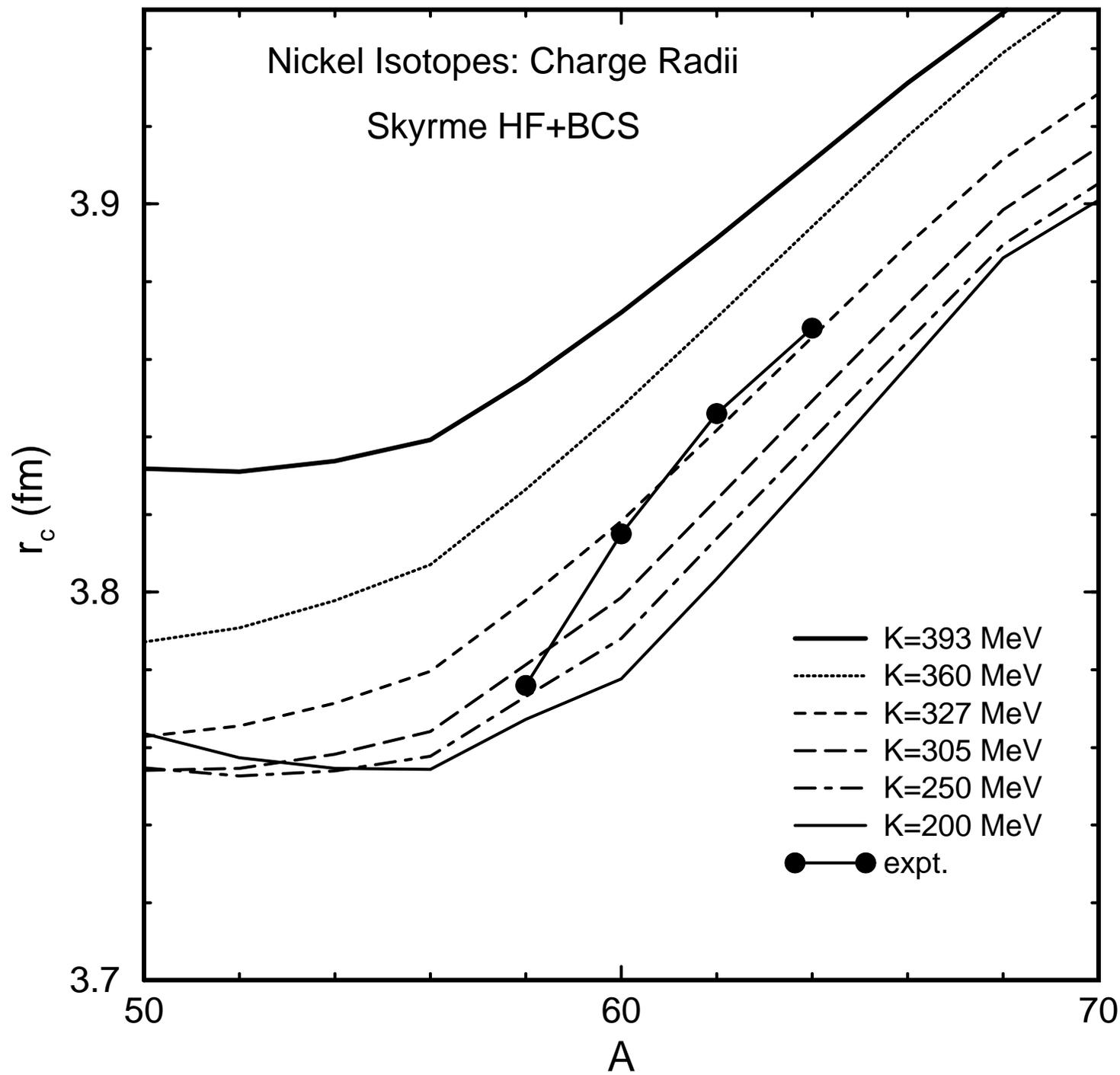
\item[Fig. 1.] {Charge radii for Ni isotopes with various Skyrme
    forces. The experimental values for a few nuclei from
Refs.$^{17,18)}$ are shown by solid dots.}

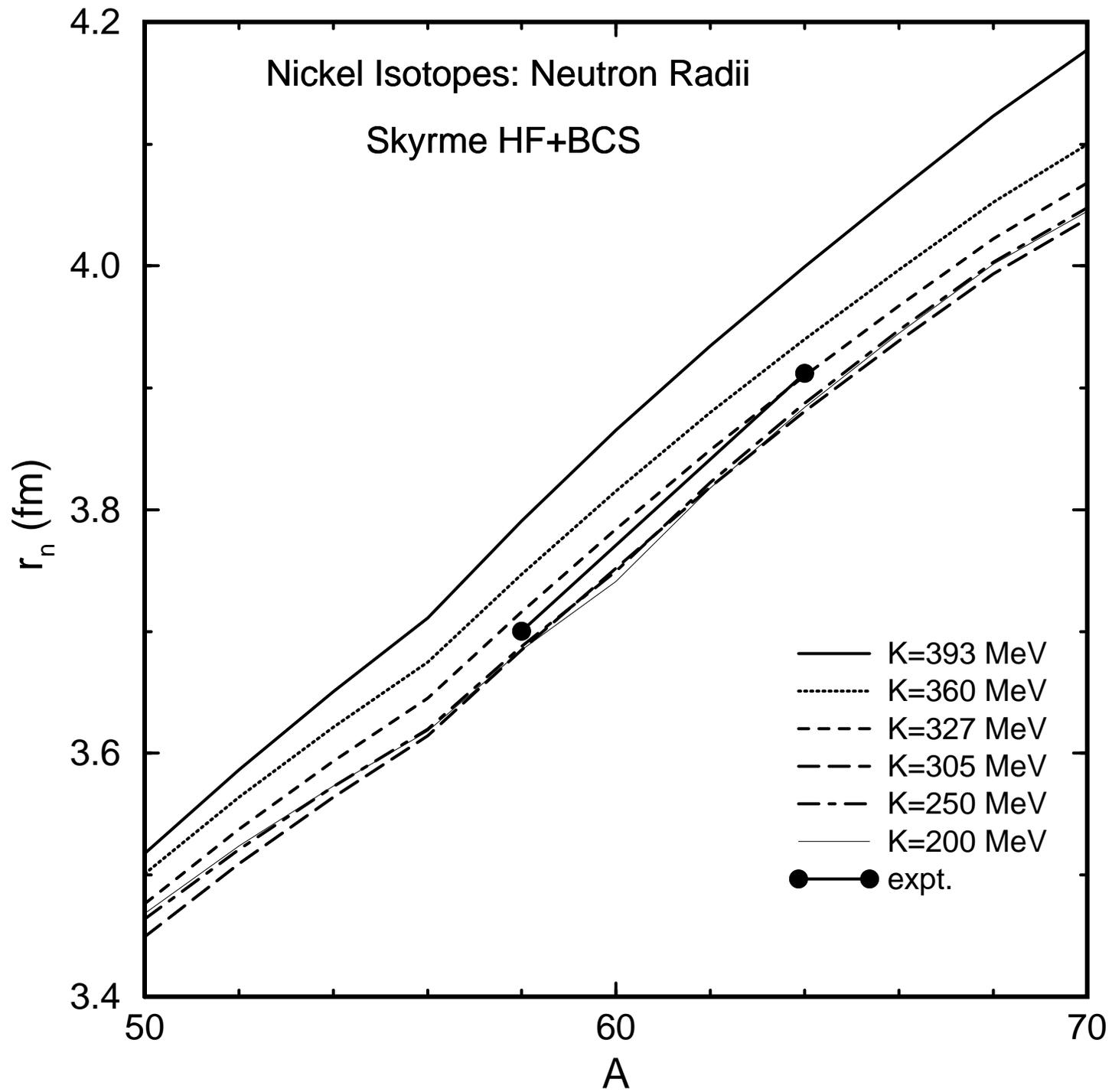
\item[Fig. 2.] {Neutron radii for Ni isotopes with various Skyrme
    forces. The experimental values for $^{58}$Ni and $^{64}$Ni
    from Refs.$^{19)}$ are shown by solid dots.}

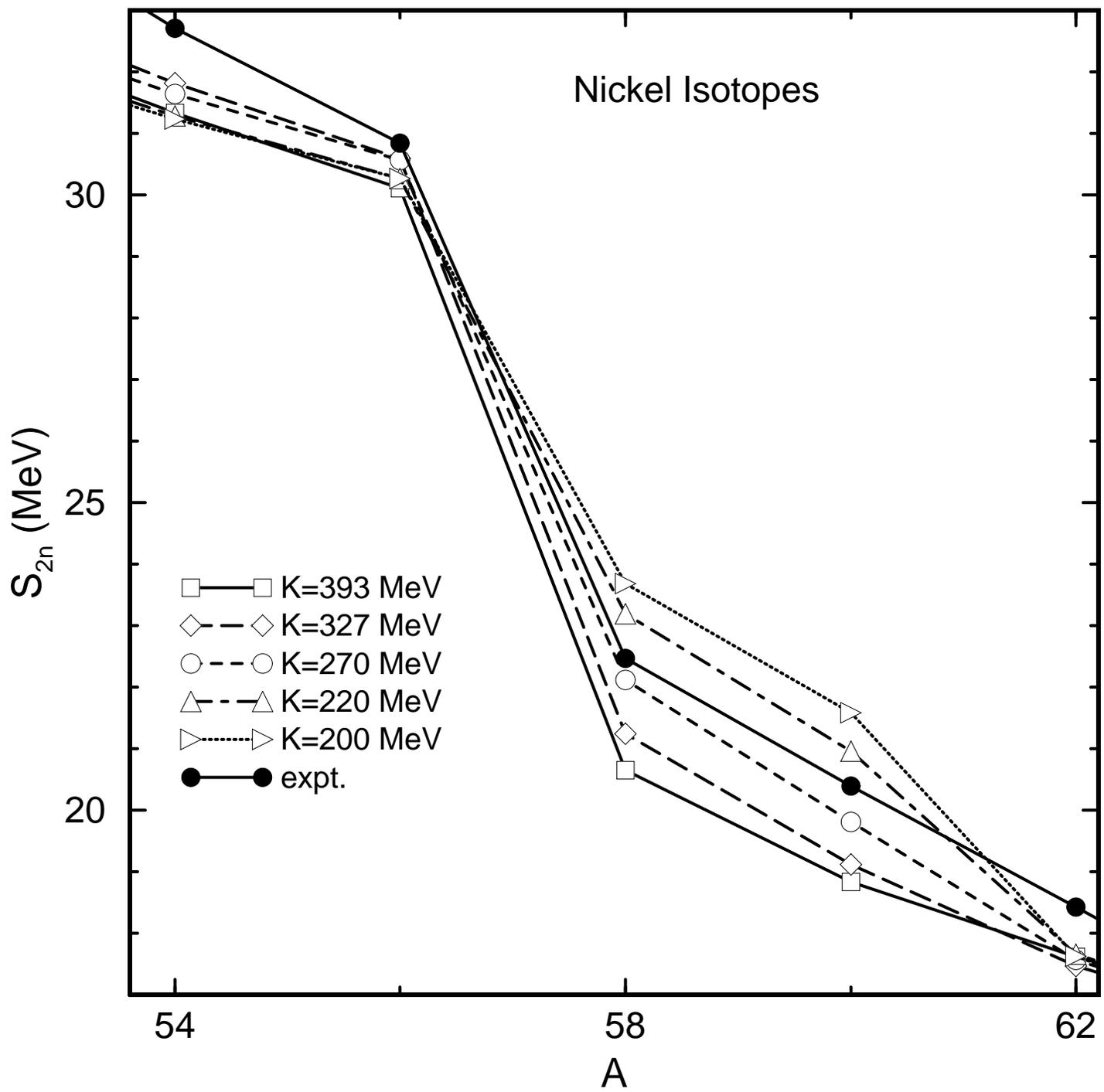
\item[Fig. 3.] {Two-neutron separation energy S$_{2n}$ for Ni isotopes
obtained with various Skyrme forces. Comparison with the experimental
data from Ref.$^{20)}$ is also made.}

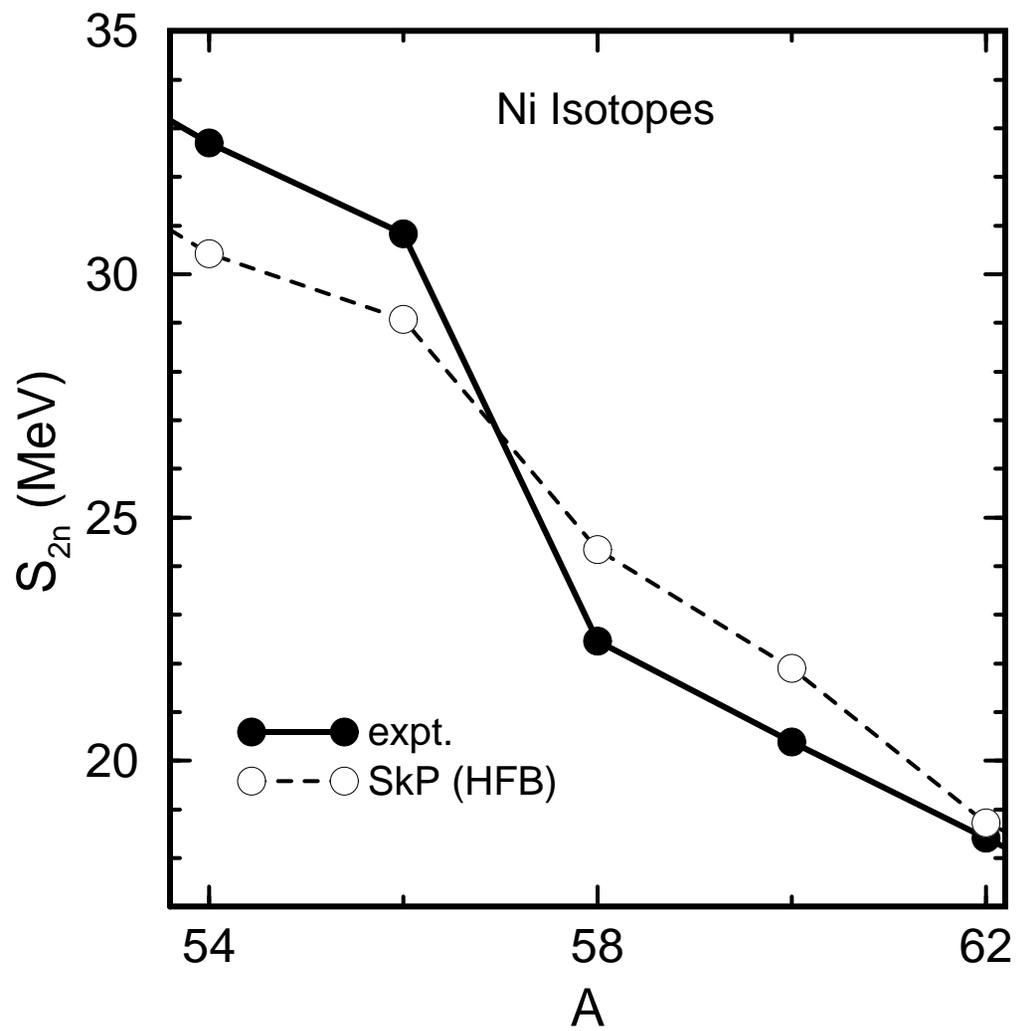
\item[Fig. 4.] {Total binding energy of Zr isotopes near the neutron
drip line calculated with the Skyrme forces.}

\end{description}



\begin{table}
\begin{center}
\caption{The nuclear matter properties of the various Skyrme 
interactions. The saturation energy $E/A$ has been fixed at $-$16.0 MeV 
and the effective mass m* has been kept at 0.79.}
\bigskip
\begin{tabular}{c c c c c c c}
\hline
& Force && $\rho_0$ ($fm^{-3}$) & K (MeV) & $a_{sym}$ (MeV) & \\
\hline
 & I  && 0.140 & 393 & 38.0 & \\
 & II && 0.145 & 360 & 32.4 &\\ 
 & III && 0.150 & 327 & 32.5 &\\
 & IV && 0.155 & 305 & 32.3 & \\
 & V  && 0.160 & 249 & 32.5 & \\
 & VI && 0.165 & 220  & 32.0 &\\
 & VII && 0.170 & 200 & 31.8 & \\
\hline
\end{tabular}
\end{center}
\end{table}


\begin{table}
\begin{center}
\caption{The total binding energy (MeV) of nuclei obtained from 
the HF+BCS calculations with the Skyrme interactions. The empirical 
values (exp.) are shown for comparison.}
\bigskip
\begin{tabular}{c c c c c c  c c c c c c c}
\hline
 &        &&      &&     &     &     & K (MeV) &  &  &  & \\
 &Nucleus && exp. && 393 & 360 & 327 & 305 & 249 & 220 & 200 &\\
\hline
 &~~$^{16}$O  && $-$127.6 && $-$127.5 & $-$127.5 & $-$125.5 & $-$127.5 & 
$-$127.5 &  $-$127.6 & $-$127.7 &\\
 &~~$^{40}$Ca && $-$342.0 && $-$342.5 & $-$342.4 & $-$342.2 & $-$342.0 & 
$-$342.4 &  $-$342.5 & $-$342.4 &\\
 &~~$^{90}$Zr && $-$783.9 && $-$786.6 & $-$786.0 & $-$785.6 & $-$785.2 &
 $-$785.3 &  $-$784.6 & $-$784.3 &\\
 &~~$^{116}$Sn && $-$988.7 && $-$989.2 & $-$988.1 & $-$987.0 & $-$986.1 &
 $-$986.1 &  $-$985.5 & $-$985.4 &\\
 &~~$^{120}$Sn && $-$1020.5 && $-$1021.3 & $-$1020.1 & $-$1019.6 & $-$1019.3
 &   $-$1019.1 &  $-$1018.9 & $-$1018.9 & \\
 &~~$^{124}$Sn && $-$1050.0 && $-$1051.0 & $-$1049.7 & $-$1050.0 & $-$1050.5 & 
$-$1049.9 &  $-$1050.1 & $-$1050.3 & \\
&~~$^{208}$Pb && $-$1636.5 && $-$1636.5 & $-$1635.7 & $-$1636.0 & $-$1636.5 & 
$-$1636.2 &  $-$1637.2 & $-$1637.0 & \\
\hline

\end{tabular}
\end{center}
\end{table}

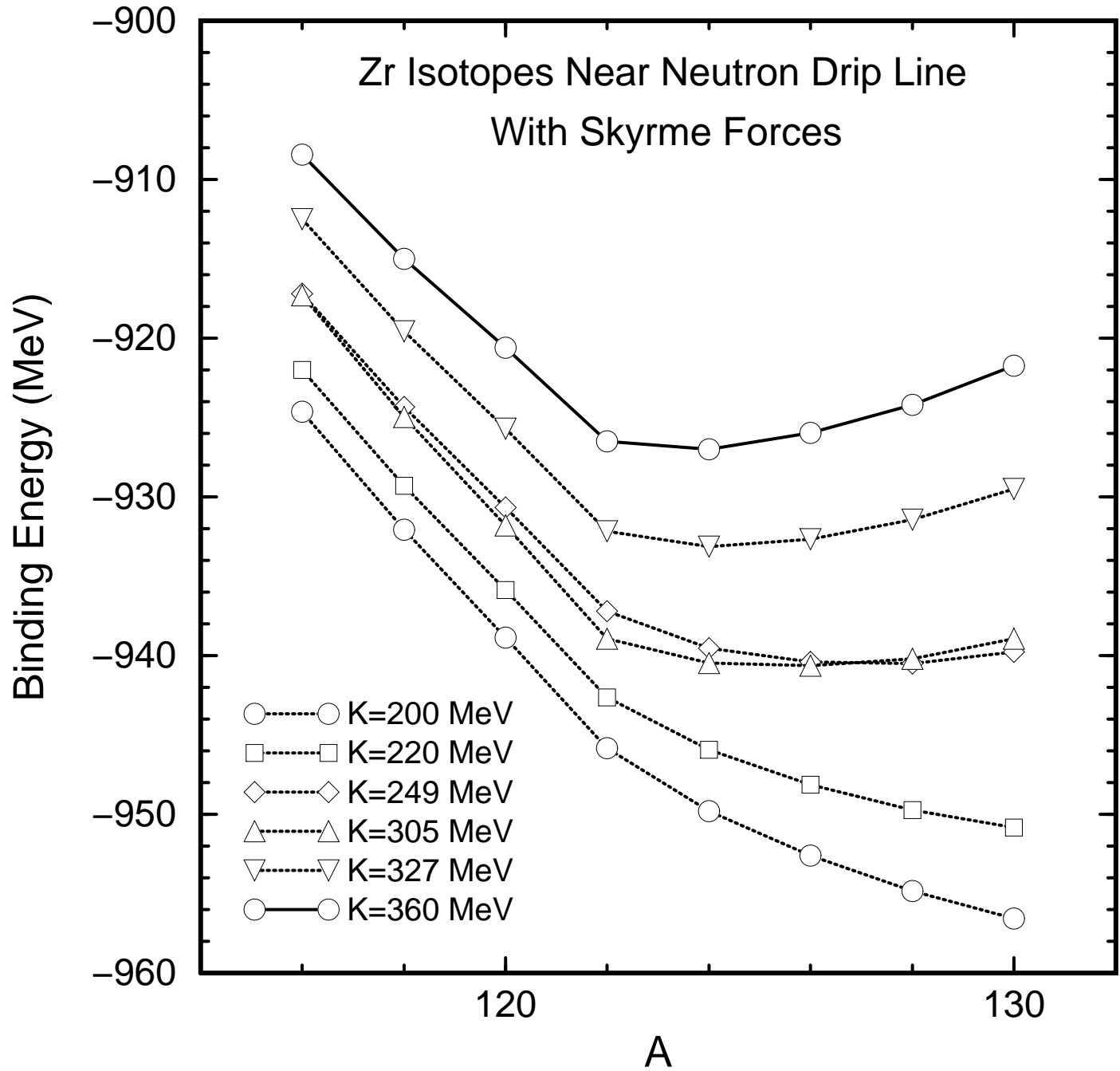
\end{document}